\documentclass[journal]{IEEEtran}
\usepackage{mathrsfs}
\usepackage{tabu}
\usepackage{bbm}
\usepackage{amsmath,amscd,amssymb,latexsym}
\usepackage[all]{xy}
\usepackage{pgf,tikz}

\newtheorem{theorem}{Theorem}[section]
\newtheorem{lemma}[theorem]{Lemma}

\newtheorem{example}[theorem]{Example}

\newtheorem{remark}[theorem]{Remark}

\ifCLASSINFOpdf
\else
\fi
\begin{document}
%
\title{ New quaternary codes derived  from posets of the
disjoint union of two chains }
%
%
%

\author{Xiaomeng Zhu,~
     Yansheng Wu~ and Qin Yue
\thanks{Manuscript received October 09, 2019; accepted October 27, 2019. The paper was supported by the National Natural Science Foundation of China under Grant 61772015 and the  Foundation of Science and Technology on Information Assurance Laboratory under Grant KJ-17-010. The associate editor coordinating the review of this letter and approving it for publication was Bal\'azs Matuz. (Corresponding author: Qin Yue.)}

\thanks{X. Zhu and Q. Yue are with the Department of Mathematics, Nanjing University of Aeronautics and Astronautics,
Nanjing, 211100, P. R. China and State Key Laboratory of Cryptology, P. O. Box 5159, Beijing, 100878, P. R. China. (e-mail: mooneernanjing@163.com; yueqin@nuaa.edu.cn).}

\thanks{Y. Wu is with the Department of Mathematics, Nanjing University of Aeronautics and Astronautics,
Nanjing, 211100, P. R. China. The author is now with the Department of Mathematics, Ewha Womans University, Seoul, 03760, South Korea. (e-mail: wysasd@163.com).}

\thanks{Digital Object Identifier ~~~~~~~~~}
}
\maketitle
\begin{abstract} Based on the generic construction of linear codes, we construct  linear codes over the ring $\Bbb Z_4$  via posets of the
disjoint union of two chains.  We determine the Lee weight distributions of the quaternary codes. Moreover, we  obtain some new linear quaternary codes and good  binary  codes by using the Gray map.

\end{abstract}


%
\IEEEpeerreviewmaketitle

\section{Introduction}
%
%
%
%
\IEEEPARstart{T}{he} research of codes over finite rings can be traced back to the early 1970s.  In a landmark paper \cite{H}, the authors
showed that certain nonlinear binary codes with excellent error-correcting capabilities
can be identified as images of linear codes over $\Bbb Z_4$ under the Gray map. This motivated
the study of codes over finite rings. Since that, a great deal of attention
has been given to codes over finite rings in the middle of 1990s because of their
new role in algebraic coding theory and  their  useful applications. Due to the relation to  lattices, designs, and low correlation
sequences, codes over $\Bbb Z_4$ remain a special topic
of interest in the field  \cite{W}.

{Recently},  Chang and Hyun  \cite{CH} constructed some infinite families of  binary optimal and minimal  linear codes  via simplicial complexes. Then Wu, Hyun, and Yue \cite{WHY} extended the construction of linear codes to posets.

 Inspired by their work, we focus on the construction of quaternary codes via  the posets of the
disjoint union of two chains in this short letter.  Let $n$ be a positive integer and $L$ be a subset of $\Bbb Z_4^n \backslash \{{\bf 0}\}$.  We define a linear code $\mathcal {C}_{L}$  over the ring $\Bbb Z_4$ as follows:
 \begin{equation}\mathcal {C}_{L}=\{c_{\mathbf{a}}=(\langle \mathbf{a}, \mathbf{l}\rangle )_{\mathbf{l}\in L}|\mathbf{a}\in \Bbb Z_4^n\},\end{equation}
where $\langle \cdot, \cdot \rangle$ is the Euclidean inner product on $\Bbb Z_4$.

The rest of this paper is organized as follows.  Section II introduces basic concepts on posets, Lee weight, and the Gray map.  Sections III determines the Lee weight distribution of the quaternary code $\mathcal {C}_{L}$ in Equation (1).  Based on the results, we  find some new quaternary codes, obtain some good binary codes, and  present some examples  in Section IV.   Section V concludes the letter.

\section{Preliminaries}

\subsection{Posets  of the disjoint union of two chains }

A subset $S$ of the Cartesian product $X\times X$ is called a partial order on the set $X$ if the following conditions are satisfied for every three elements $x,y$, and $z$ of $X$:

$(i)$ $(x,x)\in S$. 

$(ii)$ If $(x,y)\in S$ and $(y,x)\in S$, then $x=y$. 

$(iii)$ If $(x,y)\in S$ and $(y,z)\in S$, then $(x,z)\in S$. 

For convenience, we use $x\preceq y$ to denote $(x,y)\in S$, which means $x\prec y$ or $x=y$. An ordered pari $(X, \preceq)$ is called a poset if $\preceq $ is a partial order on the set $X$. Two distinct elements $x$ and $y$ in a poset $(X,\preceq)$ are called comparable if either $x\preceq y$ or $y\preceq x$, and incomparable otherwise. A poset $(X,\preceq)$ can be represented by Hasse diagram. We draw a Hasse diagram for the poset by representing each element of $X$ by a distinct point so that whenever $x\prec y$, {\bf the point representing $y$ is situated higher than the point representing $x$}. For more details, the reader is referred to \cite{N}.

Let $(\mathbb{P},\preceq)$ be a poset on $[n]$, where  $n$ is a positive integer and $[n]=\{1,2,\cdots,n\}$.  
A non-empty subset $I$ of $\mathbb{P}$ is called an {\bf order ideal} if $x \in I $ and $y \preceq x$ imply $y \in I$.   For a subset $E$ of $\mathbb{P}$, the smallest order ideal of $\mathbb{P}$ containing $E$ is denoted by $\langle E\rangle$. For an order ideal $I$ of $\mathbb{P}$,  we use $I(\mathbb{P})$ to denote the set of order ideals contained in $I$ of $\mathbb{P}$.

Let $\Bbb F_2$ be the finite field with  two elements. Assume that $n$ is a positive integer.
For a vector ${\bf v}=(v_1,\ldots, v_n)\in \Bbb F_2^n$, we define the support of ${\bf v}$ as follows: $\mbox{supp}({\bf v})=\{i: v_i\ne 0\}$, so
 the Hamming weight $wt({\bf v})$ of ${\bf v}\in \mathbb{F}^n_2$ is equal to  the size of $\mathrm{supp}({\bf v})$.
 Let $2^{[n]}$ denote the power set of $[n]$, then there is  a bijection: $\Bbb F_2^n\rightarrow 2^{[n]}, {\bf v}\mapsto \mbox {supp}({\bf v})$.  {\bf Throughout this letter, we will identify a vector in $\mathbb{F}_2^n$  with its support.} 
 
 Let $X$ be a subset of $\mathbb{F}_2^n$.  Chang and Hyun \cite{CH} introduced the following $n$-variable generating function associated with the set $X$:
$$\mathcal{H}_{X}(x_1,x_2\ldots, x_n)=\sum_{{\bf u}\in X}\prod_{i=1}^mx_i^{u_i}\in \mathbb{Z}[x_1,x_2, \ldots, x_n],
$$
where ${\bf u}=(u_1,u_2,\ldots, u_n)\in \mathbb{F}_2^n$ and $\mathbb{Z}$ is the ring of integers. 


 
 For $m\in [n]$, the authors \cite{WHY} defined the posets of disjoint union of two chains, denoted by $\mathbb{P}=(m\oplus n, \preceq)$ and  obtained  
some basic lemmas on  posets of disjoint union of two chains. 
The  Hasse diagram of the posets is given  in Figure 1.
\[
\begin{tikzpicture}
\draw (0, -1)[fill = black] circle (0.05);
\draw (0, -2)[fill = black] circle (0.05);
\draw (0, -2.5)[fill = black] circle (0.05);
\draw (0, -3)[fill = black] circle (0.05);
\draw (0, -3.5)[fill = black] circle (0.05);
\draw (0, -4)[fill = black] circle (0.05);
\draw (0, -5)[fill = black] circle (0.05);

\draw (3, -1)[fill = black] circle (0.05);
\draw (3, -2)[fill = black] circle (0.05);
\draw (3, -2.5)[fill = black] circle (0.05);
\draw (3, -3)[fill = black] circle (0.05);
\draw (3, -3.5)[fill = black] circle (0.05);
\draw (3, -4)[fill = black] circle (0.05);
\draw (3, -5)[fill = black] circle (0.05);

\draw[thin] (0, -5) --(0,-4);

\draw[thin] (0,-2)--(0,-1);
\draw[thin] (3, -5) --(3,-4);

\draw[thin] (3,-2)--(3,-1);

\node  at (0, -1) [left] {$~~~m~~$};
\node  at (0, -2) [left] {$~~~m-1~~$};
\node  at (0, -4) [left] {$~~~2~~$};
\node  at (0, -5) [left] {$~~~1~~$};
\node  at (3, -4) [right] {$~~m+2~~$};
\node  at (3, -5) [right] {$~~m+1~~$};
\node  at (3, -1) [right] {$~~~n~~$};\node  at (3, -2) [right] {$~~~n-1~~$};

\node  at (2, -5.5) [below] {Figure 1 $\mathbb{P}=$($m\oplus n, \preceq$)};

\end{tikzpicture}
\]

\begin{lemma}{\rm \cite[Lemma 2.1]{WHY} Let $i$ and $j$ be two integers with $1\le i\le m$ and  $m+1\le j\le n$. An order ideal of $\mathbb{P}=(m\oplus n, \preceq)$ must be one of the following forms:
 $$[i]; ~[j]\backslash [m]; ~ [i] \cup ([j]\backslash [m]).$$}
\end{lemma}

\begin{lemma}{\rm \cite[Lemma 2.2]{WHY} Suppose that $I$ is an order ideal  of $\mathbb{P}=(m\oplus n, \preceq)$.

$(1)$ If $I=[i]$, then $$\mathcal{H}_{I(\mathbb{P})}(x_1,x_2\ldots, x_n)=1+\sum_{k=1}^ix_1\cdots x_k.$$

$(2)$ If $I=[j]\backslash [m]$, then $$\mathcal{H}_{I(\mathbb{P})}(x_1,x_2\ldots, x_n)=1+\sum_{l=m+1}^jx_{m+1}\cdots x_l.$$

$(3)$  If $I=[i] \cup ([j]\backslash [m])$, then \begin{eqnarray*}&&\mathcal{H}_{I(\mathbb{P})}(x_1,x_2\ldots, x_n)
=1+\sum_{k=1}^ix_1\cdots x_k\\
&+&\sum_{l=m+1}^jx_{m+1}\cdots x_l+\sum_{k=1}^i \sum_{l=m+1}^j x_1\cdots x_kx_{m+1}\cdots x_l .\end{eqnarray*}}
\end{lemma}

We give an example to illustrate the meaning of notations and Lemmas 2.1 and 2.2.

  \begin{example} Let $n=6$, $m=4$, and $\mathbb{P}=(4\oplus 6, \preceq)$.

  $(1)$ If $I_1=[4]=\{1,2,3,4\}$, then  $I_1(\mathbb{P})=\big\{\emptyset, \{1\}, \{1,2\}, \{1,2,3\}, \{1,2,3,4\}\big\}$ and $$\mathcal{H}_{I_1(\mathbb{P})}(x_1,x_2\ldots, x_6)=1+x_1+x_1x_2+x_1x_2x_3+x_1x_2x_3x_4.$$

  $(2)$ If $I_2=\{5,6\}$, then  $I_2(\mathbb{P})=\big\{\emptyset, \{5\}, \{5,6\}\big\}$ and  $$ \mbox{     }\mathcal{H}_{I_2(\mathbb{P})}(x_1,x_2\ldots, x_6)=1+x_5+x_5x_6.$$

  $(3)$ If $I_3=\{1,2,5\}$, then  $I_3(\mathbb{P})=\big\{\emptyset, \{1\}, \{1,2\}, \{5\}, \{1,5\}, \{1,2,5\}\big\}$
  and $$\mathcal{H}_{I_3(\mathbb{P})}(x_1,x_2\ldots, x_6)=1+x_1+x_1x_2+x_5+x_1x_5+x_1x_2x_5.$$    \end{example}

\subsection{Lee weight and Gray map}

Let $\Bbb Z_4$ be the ring of integers module $4$, $n$ be a positive integer, and $\Bbb Z_4^n$ be the set of $n$-tuples over $\Bbb Z_4$. Any non-empty subset $\mathcal{C}$ of $\Bbb Z_4^n$ is called a quaternary code. A linear code $\mathcal{C}$ of length $n$ is a submodule of
$\Bbb Z_4^n$.


 Let $\mathbf{ x}=(x_1,x_2,\cdots, x_n)$ and $\mathbf{y}=(y_1,y_2, \cdots, y_n)$ be two vectors of $\Bbb Z_4^n$. The Euclidean inner product of $\mathbf{x}$ and $\mathbf{y}$ is defined by $\langle \mathbf{x},\mathbf{y} \rangle=\sum_{i=1}^nx_iy_i\in \Bbb Z_4$.


For each $u\in \Bbb Z_4$ there is an unique representation $u=a+2b$, where $a, b\in \Bbb F_2$.   The Gray map $\phi$ from $\Bbb Z_4$ to $\Bbb F_2^2$ is defined by $$\phi: \Bbb Z_4\to \Bbb F_2^2, a+2b\mapsto (b,a+b).$$
 Any vector ${\bf x}\in \Bbb Z_4^n$ can be written as $\mathbf{x}={a}+2{b}$, where ${\bf a}, {\bf b}\in \Bbb F_2^n$. The map $\phi$ can be extended naturally from $\Bbb Z_4^n$ to $\Bbb F_2^{2n}$ as follows:
 $$\phi: \Bbb Z_4^n\to \Bbb F_2^{2n},{\bf x}={\bf a}+2{\bf b}\mapsto ({\bf b}, {\bf a+b}).$$
 The Hamming weight of a vector ${a}$ of length $n$ over $\Bbb F_2$ is defined to be the number of nonzero entries in the vector ${a}$. The Lee weight of a vector $\mathbf{ x}$ of length $n$ over $\Bbb Z_4$ is defined to be the Hamming weight of its Gray image as follows:
$$w_L({\bf x})=w_L({\bf a}+2{\bf b})=w_H({\bf b})+w_H({\bf a+b}).$$
The Lee distance of ${\bf x,y}\in \Bbb Z_4^n$ is defined as $w_L({\bf x-y})$. It is easy to check that the Gray map is an isometry from $(\Bbb Z_4^n, d_L)$ to $(\Bbb F_2^{2n}, d_H)$. In general,
the Gray image of a quaternary code is not necessarily a binary linear code, since the Gray map is
not $\Bbb F_2$-linear.

 An $(n, K, d)_L$ quaternary code $\mathcal{C}$  is a subset with size $K$ of  $\Bbb Z_4^n$ with minimum  Lee  distance $d$.
 Let $A_i$ be the number of codewords in $\mathcal C$
 with Lee weight $i$. Suppose that the set $\{i_1,\cdots, i_k\}$ are all  Lee weights of codewords of a quaternary code $\mathcal{C}$.
 The Lee  weight enumerator of
 $\mathcal C$ is defined by
 $A_{i_1}z^{i_1}+\cdots+A_{i_k}z^{i_k}.$
The sequence $( A_{i_1},  \ldots, A_{i_k})$ is called the Lee weight distribution of
 $\mathcal C$. A code $\mathcal{C}$ is $t$-weight if the number of   nonzero elements in the set $\{i_1,\cdots, i_k\}$ is equal to $t$.

\section{ Lee weight distributions of quaternary codes}





Let $I$ be an order ideal in $\mathbb{P}=(m\oplus n, \preceq)$, which is introduced in Section II.  Let $D=(I(\mathbb{P}))^c=\Bbb F_2^n \backslash I(\mathbb{P})$ and $ L=D+2\Bbb F_2^n\subset \Bbb Z_4^n$. In this section, we will determine the Lee weight distribution of the  code $\mathcal{C}_L$ in Equation (1).

First of all, from Equation (1), it is easy to check that  the code $\mathcal{C}_L$ is a linear quaternary code. The length of the code $\mathcal{C}_{L}$ is $|L|$. Its size is at most $4^n$.

 Note that in Equation (1) if $\mathbf{a}=\mathbf{0}$, then $w_L(c_{\mathbf{a}})=0$. Next we  assume that $\mathbf{a}\neq \mathbf{0}$.
Suppose that $\mathbf{a}=\boldsymbol{\alpha}+2\boldsymbol{\beta}$, $\mathbf{l}={\bf t_1}+2{\bf t_2}$, where ${\boldsymbol{\alpha}=(\alpha_1, \cdots, \alpha_n),\boldsymbol{\beta}=(\beta_1, \cdots, \beta_n)}\in \Bbb F_2^n$, ${\bf t_1}\in D$, and ${\bf t_2}\in \Bbb F_2^n$.   Then the Lee weight of the codeword $c_{\mathbf{a}}$ of the code $\mathcal{C}_{L}$ becomes that
\begin{eqnarray}&&w_L(c_{\mathbf{a}})\nonumber\\
&=&w_L((\langle \boldsymbol{\alpha}+2\boldsymbol{\beta},{\bf t_1}+2{\bf t_2}\rangle ) _{{\bf t_1}\in D, {\bf t_2}\in \Bbb F_2^n})\nonumber\\
&=&w_L((\boldsymbol{\alpha}{\bf t_1}+2(\boldsymbol{\alpha}{\bf t_2}+\boldsymbol{\beta}{\bf t_1}))_{{\bf t_1}\in D, {\bf t_2}\in \Bbb F_2^n})\nonumber\\
&=&w_H((\boldsymbol{\alpha}{\bf t_2}+\boldsymbol{\beta}{\bf t_1})_{{\bf t_1}\in D, {\bf t_2}\in \Bbb F_2^n})\nonumber\\
&+&w_H(((\boldsymbol{\alpha}+\boldsymbol{\beta}){\bf t_1}+\boldsymbol{\alpha}{\bf t_2})_{{\bf t_1}\in D, {\bf t_2}\in \Bbb F_2^n})\nonumber\\
&=&|L|-\frac12\sum_{y\in\Bbb F_2}\sum_{\bf t_1\in D}\sum_{\bf t_2\in \Bbb F_2^n}(-1)^{(\boldsymbol{\alpha}{\bf t_2}+\boldsymbol{\beta}{\bf t_1})y}\nonumber\\
&+&|L|-\frac12\sum_{y\in\Bbb F_2}\sum_{{\bf t_1}\in D}\sum_{\bf t_2\in \Bbb F_2^n}(-1)^{((\boldsymbol{\alpha}+\boldsymbol{\beta}){\bf t_1}+\boldsymbol{\alpha}{\bf t_2})y}\nonumber\\
&=&|L|-\frac12\sum_{{\bf t_1}\in D}(-1)^{\boldsymbol{\beta}{\bf t_1}}\sum_{\bf t_2\in \Bbb F_2^n}(-1)^{\boldsymbol{\alpha}{\bf t_2}}\nonumber\\
&-&\frac12\sum_{{\bf t_1}\in D}(-1)^{\bf (\boldsymbol{\alpha}+\boldsymbol{\beta})t_1}\sum_{\bf t_2\in \Bbb F_2^n}(-1)^{\boldsymbol{\alpha}{\bf t_2}}\nonumber\\
&=&|L|-2^{2n-1}\delta_{{\bf 0},\boldsymbol{\alpha}}(\delta_{{\bf 0},\boldsymbol{\beta}}+\delta_{{\bf 0},\boldsymbol{\alpha}+\boldsymbol{\beta}})\nonumber\\
&+&2^{n-1}\delta_{{\bf 0},\boldsymbol{\alpha}}\sum_{{\bf t_1}\in I(\mathbb{P})}((-1)^{\boldsymbol{\beta}{\bf t_1}}+(-1)^{(\boldsymbol{\alpha}+\boldsymbol{\beta}){\bf t_1}}),
\end{eqnarray}
where $\delta $ is the Kronecker delta function.




By Lemmas 2.1 and 2.2, we just need to consider the following case.

\begin{theorem} {\rm Let $n\ge 2$ be a positive integer.

 $(1)$ If $I=[i]$ and $1\le i\le m$, then the Lee weight distribution of  the code $\mathcal{C}_L$ in Equation (1) is presented in Table I.  \begin{table}[h] 
\caption{Lee weight distribution of the code in Theorem 3.1 (1)}  
\begin{center}  
\begin{tabu} to 0.4\textwidth{X[2,c]|X[1,c]} 
\hline 
Lee weight &Frequency\\
\hline
$0$&$1$\\
\hline
$2^n(2^n-i-1)$&$2^n(2^n-1)$\\
 \hline
$2^{n+1}(2^{n-1}+s-i)~(0\le s< i)$&$2^{n-i}{i\choose s}$\\
\hline
$2^{2n}$&$2^{n-i}-1$
\\
\hline
\end{tabu}  
\end{center}  
\end{table}

$(2)$  If $I=[j]\backslash [m]$ and $m+1\le j\le n$, then the Lee weight distribution of  the code $\mathcal{C}_L$ in Equation (1) is presented in Table II.
\begin{table} [h] 
\caption{Lee weight distribution of the code in Theorem 3.1 (2)}  
\begin{center}  
\begin{tabu} to 0.5\textwidth{X[2,c]|X[1,c]} 
\hline 
Lee weight &Frequency\\
\hline
$0$&$1$\\
 \hline
 $2^n(2^n-(j-m)-1)$&$2^n(2^n-1)$\\
 \hline
$2^{n+1}(2^{n-1}+t-(j-m))~ (0\le t< j-m)$&$2^{n-(j-m)}{j-m\choose t}$\\
\hline
$2^{2n}$&$2^{n-(j-m)}-1$
\\
\hline
\end{tabu}  
\end{center}  
\end{table}

$(3)$  If $I=[i] \cup ([j]\backslash [m])$, $1\le i\le m$, and  $m+1\le j\le n$, then the Lee weight distribution of the code $\mathcal{C}_L$ in Equation (1) is presented in Table III. 
\begin{table} [h] 
\caption{Lee weight distribution of the code in Theorem 3.1 (3)}  
\begin{center}  
\begin{tabu} to 0.5\textwidth{X[2,c]|X[1,c]} 
\hline 
Lee weight &Frequency\\
\hline
$0$&$1$\\
 \hline
 $2^n(2^n+m-i-j-1-i(j-m))$&$2^n(2^n-1)$\\
 \hline
 $2^{n+1}(2^{n-1}+s+t+2st-(s+1)(j-m)-(t+1)i) $&\\
 $0\le s\le i, 0\le t\le (j-m),$&$2^{n-i-(j-m)}{i\choose s}{j-m\choose t}$\\
 $(s,t)\neq (i, j-m)$&\\
\hline
$2^{2n}$&$2^{n-i-(j-m)}-1$
\\
\hline
\end{tabu}  
\end{center}  
\end{table}

}

\end{theorem}

{\bf Proof} (1) It is easy to check that the length of the quaternary code $\mathcal{C}_L$ is $2^n(2^n-i-1)$.  By   Equation (2), if $\alpha\neq 0$, then $w_L(c_{\mathbf{a}})=|L|$; if $\alpha=0$ and $\beta\neq 0$, then \begin{equation}w_L(c_{\mathbf{a}})=|L|+2^n\sum_{{\bf t_1}\in I(\mathbb{P})}(-1)^{\beta{\bf t_1}}.
\end{equation}
By the definition of generating function in Section II,
\begin{equation}w_L(c_{\mathbf{a}})=|L|+2^n\mathcal{H}_{I(\mathbb{P})}((-1)^{\beta_1},\ldots, (-1)^{\beta_n}).
\end{equation}
Thanks to \cite[Theorem 3.1 (1)]{WHY}, we have
 $$\mathcal{H}_{I(\mathbb{P})}((-1)^{\beta_1},\ldots, (-1)^{\beta_n})=1+2s-i,$$
where $s$ is the number of such $k$ with $1\le k \le i$ satisfying $$w_H((\beta_1, \beta_2, \cdots, \beta_k))\equiv 0 \pmod 2.$$
Note that here $s=i$ means $\beta_1=\cdots=\beta_i=0$.
The frequency of each codeword  should be computed by the vector ${\bf a}$.

(2) It is easy to check that the length of the quaternary code $\mathcal{C}_L$ is $2^n(2^n+m-j-1)$.  Thanks to \cite[Theorem 3.1 (2)]{WHY}, we have  $$\mathcal{H}_{I(\mathbb{P})}((-1)^{\beta_1},\ldots, (-1)^{\beta_n})=1+2t-(j-m),$$
where $t$ is the number of such $l$ with $m+1\le l \le j$ satisfying $$w_H((\beta_{m+1}, \beta_{m+2}, \cdots, \beta_l))\equiv 0 \pmod 2.$$
By Equations (2) and (4), the results hold and the frequency of each codeword of the codes should be computed by the vector ${\bf a}$.

(3) It is easy to check that the length of the quaternary code $\mathcal{C}_L$ is $2^n(2^n+m-i-j-1-i(j-m))$.  Thanks to \cite[Theorem 3.1 (3)]{WHY}, we have \begin{eqnarray*}&&\mathcal{H}_{I(\mathbb{P})}((-1)^{\beta_1},\ldots, (-1)^{\beta_n})\\
&=&1+2s-i+2t-(j-m)+(2s-i)(2t-(j-m)),\end{eqnarray*} where $s$ is the number of such $k$ with $1\le k \le i$ satisfying $$w_H((\beta_1, \beta_2, \cdots, \beta_k))\equiv 0 \pmod 2$$ and $t$ is the number of such $l$ with $m+1\le l \le j$ satisfying $$w_H((\beta_{m+1}, \beta_{m+2}, \cdots, \beta_l))\equiv 0 \pmod 2.$$ By Equations (2) and (4), the results hold and the frequency of each codeword  should be computed by the vector ${\bf a}$.

This completes the proof.
$\blacksquare$

\begin{remark} {\rm The size of the code $\mathcal{C}_L$ is $4^n$ except the case of $n=i=2$ in Theorem 3.1 (1).    When $n=i=2$ in Theorem 3.1 (1), the size of the quaternary code $\mathcal{C}_L$ is  $8$ because in this case the frequency of codewords with zero Lee weight  is $2$.}
\end{remark}

\section{New quaternary codes and optimal binary codes }

In this section, based on the results in Section III, we will present some new quaternary codes and optimal binary codes.

\begin{example}{\rm In Theorem 3.1 (1), let $n=m=i=2$. Then $D=\{(0,1)\}$ and $L=\{(0,1),(0,3), (2,1), (2,3)\}$. According  to  \cite{AA},   $\mathcal{C}_L$ is a new two-weight quaternary linear code with parameters $(4, 8, 4)_L$. Its
Lee weight enumerator is $1+6z^4+z^8$ and its codewords are given by
\begin{eqnarray*}\big\{(0,0,0,0), (0,0,2,2), (2,2,0,0), (2,2,2,2), \\(1,3,3,1), (3,1,3,1), (1,3,3,1), (3,1,1,3)\big\}.\end{eqnarray*} }
\end{example}

\begin{example}{\rm  In Theorem 3.1 (1), let $n=3$ and $i=1$. Then $$D=\{(0,1,0), (0,0,1), (0,1,1), (1,0,1), (1, 1,0), (1,1,1)\}$$ and
\begin{eqnarray*}L&=&\big\{ (0,1,0),(0,1,2), (0,3,0),(0,3,2), (2,1,0), (2,3,0),\\
&& (2,1,2), (2,3,2), (0,0,1),(0,0,3), (0,2,1),(0,2,3), \\
&&(2,0,1), (2,2,1),(2,0,3), (2,2,3), (0,1,1),(0,1,3),\\
&& (0,3,1),(0,3,3), (2,1,1), (2,3,1),(2,1,3), (2,3,3), \\
&&(1,0,1),(1,0,3), (1,2,1),(1,2,3), (3,0,1), (3,2,1),\\
&& (3,0,3), (3,2,3) (1,1,0),(1,1,2), (1,3,0),(1,3,2), \\
&&(3,1,0), (3,3,0),(3,1,2), (3,3,2),(1,1,1),(1,1,3), \\
&&(1,3,1),(1,3,3), (3,1,1), (3,3,1),(3,1,3), (3,3,3)\big\}.
\end{eqnarray*}
According  to  \cite{AA},   $\mathcal{C}_L$ is a new two-weight quaternary linear code with parameters $(48, 64, 48)_L$. Its Lee weight enumerator is $1+60z^{48}+3z^{64}$. Note that, in the table \cite{AA}, there is only a nonlinear quaternary code with parameters $(48, 64, 48)_L$. }
\end{example}

\begin{example}{\rm In Theorem 3.1 (1), let $n=3$ and $i=3$. Then $D=\{(0,1,0), (0,0,1), (0,1,1), (1,0,1)\}$ and \begin{eqnarray*}L&=&\{ (0,1,0),(0,1,2), (0,3,0),(0,3,2), (2,1,0), (2,3,0),\\
&& (2,1,2), (2,3,2), (0,0,1),(0,0,3), (0,2,1),(0,2,3), \\
&&(2,0,1), (2,2,1),(2,0,3), (2,2,3), (0,1,1),(0,1,3),\\
&&  (0,3,1),(0,3,3), (2,1,1), (2,3,1),(2,1,3), (2,3,3),\\
&&(1,0,1),(1,0,3), (1,2,1),(1,2,3), (3,0,1), (3,2,1),\\
&&(3,0,3), (3,2,3)\}.
\end{eqnarray*} According  to  \cite{AA},   $\mathcal{C}_L$ is a new three-weight quaternary linear code with parameters $(32, 64, 16)_L$. Its Lee weight enumerator is $1+z^{16}+59z^{32}+3z^{48}$.}
\end{example}

\begin{example}{\rm In Theorem 3.1 (3), let $n=3$, $m=i=1$, and $j=2$.
Then $D=\{(0,0,1), (0,1,1), (1,0,1),  (1,1,1)\}$ and
\begin{eqnarray*}L&=&\{  (0,0,1),(0,0,3), (0,2,1),(0,2,3), (2,0,1), \\
&& (2,2,1),(2,0,3), (2,2,3),(0,1,1),(0,1,3), \\
&& (0,3,1),(0,3,3), (2,1,1), (2,3,1),(2,1,3), \\
&& (2,3,3),(1,0,1),(1,0,3), (1,2,1),(1,2,3), \\
&&(3,0,1), (3,2,1),(3,0,3), (3,2,3)(1,1,1),(1,1,3),\\
&& (1,3,1),(1,3,3), (3,1,1), (3,3,1),(3,1,3), (3,3,3)\}.
\end{eqnarray*} According  to  \cite{AA},   $\mathcal{C}_L$ is a new two-weight quaternary linear code with parameters $(32, 64, 32)_L$. Its Lee weight enumerator is $1+62z^{32}+z^{64}$. Note that, in the table \cite{AA}, there is only a nonlinear quaternary code with parameters $(32, 64, 32)_L$. }


\end{example}

Next, we will present some binary optimal codes. Recall that the Gray map $\phi$  in Section II is an isometry from $(R^m, d_L)$ to $(\Bbb F_2^{2m}, d_H)$. Then we should consider the binary codes $\phi(\mathcal{C}_{L})$.  The following lemma is needed to characterize when the Gray image of a quaternary code is linear.

\begin{lemma}{\rm \cite[Corollary 3.17] {W}
Let $\mathcal{C}$ be a  quaternary linear code, ${\bf x}_1, \cdots, {\bf x}_m$ be a set of generators
of $\mathcal{C}$, and $C = \phi(\mathcal{C})$. Then $C$ is linear if and only if $$2\alpha({\bf x}_i) \ast \alpha({\bf x}_j) \in \mathcal{C}$$ for all
$i, j$ satisfying $1 \le i \le j \le m$, where $\alpha: \Bbb Z_4 \to \Bbb F_2$ is a group homomorphism and $\alpha(0)=0,\alpha(1)=1, \alpha(2)=0,$ and $ \alpha(3)=1$, and $\ast$ denotes the componentwise product
of two vectors.}

\end{lemma}

The following are some  examples.

\begin{example} {\rm  (Continued from Example 4.1)  It is easy to check that $\{c_{(1,0)},c_{(0,1)}\}$ is the generator set of the code $\mathcal{C}_L$. By Example 4.1, we know that $$\alpha(c_{(1,0)})=\alpha(0,0,2,2)=(0,0,0,0)$$ and $$\alpha(c_{(0,1)})=\alpha(1,3,1,3)=(1,1,1,1).$$ Hence $2\alpha(c_{(1,0)}) \ast \alpha(c_{(0,1)})=(0,0,0,0) \in \mathcal{C}_L$. By Lemma 4.5,   $\phi( \mathcal{C}_L)$ is a binary linear code with parameters $[8,3, 4]$, which is optimal by \cite{G2}. }



\end{example}

\begin{example} {\rm  In Theorem 3.1 (1), let $n=2$ and $i=1$. Then $D=\{(0,1), (1,1)\}$ and $L=\{(0,1),(0,3), (1,1), (1,3), (2,1), (2,3),  (3,1), (3,3)\}$.   It is easy to check that $\{c_{(1,0)},c_{(0,1)}\}$ is the generator set of the code $\mathcal{C}_L$. Then we have that $$\alpha(c_{(1,0)})=\alpha(0,0,1,1,2,2,3,3)=(0,0,1,1,0,0,1,1)$$ and $$\alpha(c_{(0,1)})=\alpha(1,3,1,3,1,3,1,3)=(1,1,1,1, 1,1,1,1).$$
Hence  $2\alpha(c_{(1,0)}) \ast \alpha(c_{(0,1)})=(0,0,2,2,0,0,2,2) =c_{(2,0)} \in \mathcal{C}_L$. By Lemma 4.5,  $\phi(\mathcal{C}_L)$ is a binary linear code with $[16,4,8]$, which is optimal by \cite{G2}. }



\end{example}

\begin{example} {\rm  In Theorem 3.1 (1), let $n=3$ and $i=2$. Then $D=\{(0,1,0), (0,0,1), (0,1,1), (1,0,1), (1,1,1)\}$ and
\begin{eqnarray*}L&=&\{ (0,1,0),(0,1,2), (0,3,0),(0,3,2),   (2,1,0) ,(2,1,2), \\
&& (2,3,0), (2,3,2),(0,0,1),(0,0,3), (0,2,1),(0,2,3), \\
&& (2,0,1), (2,2,1),(2,0,3), (2,2,3),(0,1,1),(0,1,3), \\
&&(0,3,1),(0,3,3), (2,1,1), (2,3,1),(2,1,3), (2,3,3), \\
&&(1,0,1),(1,0,3), (1,2,1),(1,2,3), (3,0,1), (3,2,1),\\
&&(3,0,3), (3,2,3) (1,1,1),(1,1,3), (1,3,1),(1,3,3), \\
&&(3,1,1), (3,3,1),(3,1,3), (3,3,3)\}.
\end{eqnarray*}
 It is easy to check that $\{c_{(1,0,0)},c_{(0,1, 0)}, c_{(0,0,1)}\}$ is the generator set of the code $\mathcal{C}_L$. Then $$\alpha(c_{(1,0,0)})=\alpha(a, a, a, b,b)=({\bf 0}, {\bf 0},{\bf 0},{\bf 1},{\bf 1}),$$ and $$\alpha(c_{(0,1,0)})=\alpha(c,a,c,a,c)=({\bf 1},{\bf 0},{\bf 1},{\bf 0},{\bf 1}),$$ where $a=(0,0,0,0,2,2,2,2)$, $b=(1,1,1,1,3,3,3,3)$, $c=(1,1,3,3,1,1,3,3)$, ${\bf 0}=(0,0,0,0,0,0,0,0)$, and ${\bf 1}=(1,1,1,1,1,1,1,1)$.
Hence  $2\alpha(c_{(1,0,0)}) \ast \alpha(c_{(0,1,0)})=2({\bf 0}, {\bf 0}, {\bf 0}, {\bf 0} , {\bf 1}).$ Note that the Lee weight of the vector $2\alpha(c_{(1,0,0)}) \ast \alpha(c_{(0,1,0)})$ is $2\times 8=16$. By Theorem 3.1 (1), the code $\mathcal{C}_L$ has parameters $(40, 64,32)_L$. Hence  $2\alpha(c_{(1,0,0)}) \ast \alpha(c_{(0,1,0)})\notin \mathcal{C}_L$ and  $\phi(\mathcal{C}_L)$ is not a binary linear code by Lemma 4.5.
 It has a smaller minimum Hamming distance than the best known binary linear code [80,6,40] (see \cite{G2}). }


\end{example}

\begin{example}{\rm  (Continued from Example 4.2) Similarly, we have
$\alpha(c_{(1,0,0)})=({\bf 0}, {\bf 0},{\bf 0},{\bf 1},{\bf 1},{\bf 1})$ and $\alpha(c_{(0,1,0)})=({\bf 1},{\bf 0},{\bf 1},{\bf 0},{\bf 1},{\bf 1}).$ 
Hence  $2\alpha(c_{(1,0,0)}) \ast \alpha(c_{(0,1,0)})=2({\bf 0}, {\bf 0}, {\bf 0}, {\bf 0} , {\bf 1}, {\bf 1}).$ Note that the Lee weight of the vector $2\alpha(c_{(1,0,0)}) \ast \alpha(c_{(0,1,0)})$ is $2\times 16=32$. However, the code $\mathcal{C}_L$ has parameters $(48, 64,48)_L$. Hence  $$2\alpha(c_{(1,0,0)}) \ast \alpha(c_{(0,1,0)})\notin \mathcal{C}_L$$ and  $\phi(\mathcal{C}_L)$ is not a binary linear code by Lemma 4.5. It has the same minimum Hamming distance with the best known binary linear code [96, 6, 48] (see \cite{G2}).}
\end{example}

\begin{example}{\rm  (Continue of Example 4.3) Similarly, we have
$\alpha(c_{(1,0,0)})=({\bf 0}, {\bf 0},{\bf 0},{\bf 1}), \alpha(c_{(0,1,0)})=({\bf 1},{\bf 0},{\bf 1},{\bf 0}),$ and $\alpha(c_{(0,0,1)})=({\bf 0},{\bf 1},{\bf 1},{\bf 1}).$
Hence  $$2\alpha(c_{(1,0,0)}) \ast \alpha(c_{(0,1,0)})=2( {\bf 0}, {\bf 0}, {\bf 0} , {\bf 0})\in \mathcal{C}_L,$$
$$2\alpha(c_{(1,0,0)}) \ast \alpha(c_{(0,0,1)})=2( {\bf 0}, {\bf 0}, {\bf 0} , {\bf 1})=c_{(2,0,0)}\in \mathcal{C}_L, \mbox{  }$$ and
$$2\alpha(c_{(0,1,0)}) \ast \alpha(c_{(0,0,1)})=2( {\bf 0}, {\bf 0}, {\bf 1} , {\bf 0}).$$
Suppose that  $2( {\bf 0}, {\bf 0}, {\bf 1} , {\bf 0})\in \mathcal{C}_L$. Then we need find a vector ${\bf a }=(a_1,a_2, a_3)$ in $\Bbb F_2^3$ such that $\langle {\bf a, l} \rangle_{l\in L}=2( {\bf 0}, {\bf 0}, {\bf 1} , {\bf 0})$.  From $\langle {\bf a}, (0,1,0) \rangle =0$ and $\langle {\bf a}, (0,0,1) \rangle =0$, we get $a_2=a_3=0$. However in this case $\langle {\bf a}, (0,1,1) \rangle =0$, which is contradict with $\langle {\bf a, l} \rangle_{l\in L}=2( {\bf 0}, {\bf 0}, {\bf 1} , {\bf 0})$. Therefore   $\phi(\mathcal{C}_L)$ is not a binary linear code by Lemma 4.5.  It has a smaller minimum Hamming distance than the best known binary linear code [64, 6,32] (see \cite{G2}). }
\end{example}

\begin{example}{\rm (Continued from Example 4.4) Similarly, we have
$\alpha(c_{(1,0,0)})=({\bf 0}, {\bf 0},{\bf 1},{\bf 1}), \alpha(c_{(0,1,0)})=({\bf 0},{\bf 1},{\bf 0},{\bf 1}), \mbox{ and } \alpha(c_{(0,0,1)})=({\bf 1},{\bf 1},{\bf 1},{\bf 1}).$
Hence  $$2\alpha(c_{(1,0,0)}) \ast \alpha(c_{(0,1,0)})=2( {\bf 0}, {\bf 0}, {\bf 0} , {\bf 1}),$$
$$2\alpha(c_{(1,0,0)}) \ast \alpha(c_{(0,0,1)})=2( {\bf 0}, {\bf 0}, {\bf 1} , {\bf 1})=c_{(2,0,0)}\in \mathcal{C}_L, \mbox{  }$$ and
$$2\alpha(c_{(0,1,0)}) \ast \alpha(c_{(0,0,1)})=2( {\bf 0}, {\bf 1}, {\bf 0} , {\bf 1})=c_{(0,2,0)}\in \mathcal{C}_L.$$
Suppose that  $2( {\bf 0}, {\bf 0}, {\bf 0} , {\bf 1})\in \mathcal{C}_L$. Then we need find a nonzero vector ${\bf a }=(a_1,a_2, a_3)$ in $\Bbb F_2^3$ such that $\langle {\bf a, l} \rangle_{l\in L}=2( {\bf 0}, {\bf 0}, {\bf 0} , {\bf 1})$.  From $\langle {\bf a}, (0,0,1) \rangle =0$ and $\langle {\bf a}, (0,1,1) \rangle =0$, we get $a_2=a_3=0$. However in this case $\langle {\bf a}, (1,0,1) \rangle =a_1$. Namely it is 0 if and only if $a_1=0$, which is a contradiction.  Therefore  $\phi(\mathcal{C}_L)$ is not a binary linear code by Lemma 4.5. It has the same minimum Hamming distance with the best known binary linear code $[64, 6, 32]$ (see \cite{G2}).}
\end{example}





\section{Concluding remarks}
The main contributions of this letter are the following

\begin{itemize}
\item A new construction of quaternary  codes    associated with posets of the disjoint union of two chains in Equation (1).

\item The determinations  of the Lee weight distributions of  the quaternary codes (Theorem 3.1).

\item Some new quaternary codes and  good binary nonlinear codes (Tables IV and V).
\end{itemize}
By massive computation, some new quaternary codes and binary optimal codes can be
also found from Theorem 3.1 and the Gray image of the quaternary codes. As future work, it should be interesting to find more new quaternary  codes by using other posets.

\begin{table} 
\caption{Linear quaternary codes from Theorem 3.1}  
\begin{center}  
\begin{tabu} to 0.5\textwidth{X[0.8,c]|X[2.5,c]} 
\hline 
Parameters &Comparing with the $\mathbb{Z}_4$ Database\\
\hline
$(4,8,4)_L$&New\\
 \hline
 $(32,64,16)_L$&New\\
  \hline
  $(32,64,32)_L$&New (there is  only a nonlinear quaternary code with the same parameters)\\
  \hline
 $(48, 64,48)_L$&New (there is only a nonlinear quaternary code with the same parameters)\\
 \hline
\end{tabu}  
\end{center}  
\end{table}

\begin{table} 
\caption{optimal binary codes from Gray image of quaternary codes}  
\begin{center}  
\begin{tabu} to 0.5\textwidth{X[0.5,c]|X[0.5,c]|X[2.5,c]} 
\hline 
Parameters &Linearity &Comparing with some known codes\\
\hline
$[8,3,4]$&Linear &Optimal\\
 \hline
 $[16, 4,8]$&Linear&Optimal\\
 \hline
  $[64,6, 16]$&Nonlinear&The best known binary linear code $[64,6,32]$\\
\hline
 $[64,6, 32]$&Nonlinear&The best known binary linear code $[64,6,32]$\\
\hline
 $[80,6,32]$&Nonlinear&The best known binary linear code $[80,6,40]$\\
  \hline
 $[96,6, 48]$&Nonlinear&The best known binary linear code $[96,6,48]$\\
\hline

\end{tabu}  
\end{center}  
\end{table}





%

\section*{Acknowledgment}


The authors are very grateful to the reviewers and the Associate Editor for their valuable comments and suggestions to improve the quality of this paper. The authors contribute equally to this letter.

\ifCLASSOPTIONcaptionsoff
  \newpage
\fi

\end{document}